\documentclass[%
 reprint,
superscriptaddress,
 amsmath,amssymb,
 aps,
]{revtex4-2}

\usepackage{placeins}
\usepackage{float}

\usepackage{color}
\usepackage{soul}

\usepackage[compat=1.1.0]{tikz-feynman}
\usepackage{csquotes}
\usepackage{tikz}

\usepackage{graphicx}
\usepackage{dcolumn}
\usepackage{bm}

\usepackage[colorlinks,
            linkcolor=blue,
            citecolor=blue,
            urlcolor=blue]{hyperref}

\begin{document}

\preprint{APS/123-QED}

\title{Partial Relief of the Hubble Tension and a Natural \\ Self-Interacting Dark Matter Candidate From Staged Symmetry Breaking} 

\date{\today}

\author{Zachary J. Hoelscher}
\email{zachary.j.hoelscher@vanderbilt.edu} 
\affiliation{Department of Physics and Astronomy, Vanderbilt University, Nashville, TN 37235, USA}

\author{Thomas W. Kephart}
\email{thomas.w.kephart@vanderbilt.edu}
\affiliation{Department of Physics and Astronomy, Vanderbilt University, Nashville, TN 37235, USA}  

\author{Robert J. Scherrer}
\email{robert.scherrer@vanderbilt.edu}
\affiliation{Department of Physics and Astronomy, Vanderbilt University, Nashville, TN 37235, USA} 

\author{Kelly Holley-Bockelmann}
\email{k.holley@vanderbilt.edu}
\affiliation{Department of Physics and Astronomy, Vanderbilt University, Nashville, TN 37235, USA}
\affiliation{Department of Life and Physical Sciences, Fisk University, Nashville, TN 37208, USA}
\affiliation{Space Research Organisation Netherlands (SRON), Niels Bohrweg 4, Leiden, 2333 CA, The Netherlands}

\begin{abstract}
The values of the Hubble constant ($\rm{H_0}$) inferred from the cosmic microwave background (CMB) and local measurements via the distance ladder exhibit a $\sim5\sigma$ tension. In this work we propose that the tension might be partially alleviated if a subcomponent of the dark matter undergoes decays triggered by spontaneous symmetry breaking in the dark sector, where dark energy couples to the daughter species, but not the parent. This then allows interactions between dark matter and dark energy to turn on at a particular redshift, which can reduce fifth-force constraints, since early time interactions are absent. We provide an example of such an effective field theory whose structure is partially motivated by the desire for a plausible UV completion. We find that such a construction can naturally incorporate a possible self-interacting dark matter candidate with a velocity-dependent scattering cross section as a by-product of gauge invariance, though this is not necessary for alleviating the Hubble tension. This is relevant for addressing tensions between $\Lambda$CDM and small-scale structure, such as the core-cusp problem. 
\end{abstract}

\maketitle


\section{\label{sec:intro} Introduction}

The Hubble constant ($\rm{H_0}$) encodes the relation between redshift and distance from Earth; it is closely tied to the expansion rate of the Universe, originally studied by Edwin Hubble ~\cite{Hubble}. $\rm{H_0}$ can be found through either an early-time measurement or a local measurement, which produce different values. (We note, however, that some have argued that the discrepancy may be rooted in the distance ladder itself, as one-step methods may not exhibit a tension with the early-time value~\citep{Peri_2024}.) The local measurement involves employing Type Ia supernovae to determine distances to faraway galaxies, where the redshifts and distances can be used to determine $\rm{H_0}$. This method is calibrated using Cepheid variables, which are in turn calibrated using parallax measurements for distances to nearby Cepheids ~\cite{ModernAstrophysicsBook, Maguire2017, Local_H0}. 

The early-time measurement instead employs the power spectrum of the cosmic microwave background (CMB), where the Hubble constant is connected to the first peak in the CMB power spectrum ~\cite{Kamionkowski_2023}. The CMB formed at a redshift of around 1090, when the Universe cooled enough to allow electrons and protons to combine as atoms, which enabled photons to travel freely for the first time~\citep{CMB_Review}. These redshifted photons are now in the microwave band. The CMB power spectrum suggests (under the assumption of $\Lambda$CDM) a smaller $\rm{H_0}$ than the value determined by local measurement, with the two values exhibiting a $\sim5\sigma$ tension ~\cite{Local_H0, Planck_H0}. 

Many past works have proposed mechanisms to attempt to relieve the tension, though none have been fully accepted as a solution, and many face challenges in adhering to all constraints. Some have considered decays of dark matter to dark radiation ~\cite{DM_Decay, Pandey_2020}. While there is a wealth of evidence to suggest that dark matter exists ~\cite{Rubin, DMReview}, it is not known that it decays in this way. Such conversion of matter to radiation is also tightly constrained~\citep{Camarena_2016}, in part by the Integrated Sachs-Wolfe effect on the CMB~\citep{Sachs_1967, Rees_1968, DM_Decay_Constrained}. Related work has considered conversion of matter to gravitational radiation via black hole-black hole mergers, though the merger rates required to alleviate the tension are unphysically large~\citep{Hoelscher_2025}. 

Others have considered early dark energy~\citep{Karwal_2016, Poulin_2018, Poulin_2019, Valentino_2021, Gonzalez_2021, Kamionkowski_2023, Sharma_2024, Copeland_2024, Zheng_2024}, which is a hypothetical exotic field that would briefly constitute around one-tenth of the energy density of the Universe at high redshifts before redshifting away faster than radiation~\citep{Kamionkowski_2023}. This works to alleviate the Hubble tension by shrinking the sound horizon at recombination, which raises the value for $\mathrm{H_0}$ inferred from the CMB~\citep{Karwal_2016, Poulin_2019, Valentino_2021}. Some have also argued that early dark energy is a good explanation for some observations from the James Webb Space Telescope (JWST)~\citep{Shen_2024}. Many models for early dark energy exist in the literature, where one such model employs an ultralight axion field~\citep{Poulin_2018}. See~\citep{Poulin_2023_Review} for a discussion of other models. 

While it is often considered one of the more promising mechanisms for addressing the Hubble tension, early dark energy faces significant challenges, as do early-time modifications to $\Lambda$CDM in general~\citep{Vagnozzi_2021, Vagnozzi_2023}. It is notable for our work that this could necessitate new physics at late times, possibly in addition to new physics at early times, as early-time modifications alone may not be fully sufficient~\citep{Vagnozzi_2023}. For example, some have questioned whether modifications to the sound horizon are sufficient to solve the problem \cite{SoundHorizon}. It has been found that solutions that reduce $\Omega_m h^2$ conflict with observations from BAO, whereas solutions that increase $\Omega_m h^2$ conflict with observations of galaxy weak lensing \cite{SoundHorizon}. Early dark energy also faces trouble when confronted with the $S_8$ tension, which remains unresolved, even after allowing the sum of neutrino masses to vary~\citep{Reeves_2023}.

Additional works have considered modifications to gravity~\citep{Montani_2023, Schiavone_2023, Escamilla_2024, Kazem_2025}. Others have studied alleviating the tension by incorporating variation in the effective gravitational constant with redshift~\citep{Peri_2026}. Still others have considered addressing the tension within the context of braneworld cosmology, which incorporates large extra dimensions~\citep{Bag_2021}. 

In a previous work, we considered a decay chain in the dark sector, where decays were triggered by spontaneous symmetry breaking~\citep{DESI_Cascade}. This was inspired by previous work by Chen, Loeb, and Braglia, who considered generic step-like changes in the equation of state parameter for a subcomponent of the dark matter~\citep{Chen_2025, Braglia_2025}. Our work is also related to past work by Dienes on dynamical dark matter~\citep{Dienes_2012, Dienes_2012_b, Dienes_2013}. Unlike our previous work, which had four stages in its decay chain~\citep{DESI_Cascade}, we now consider a much simpler model with only two stages. This is intended to partially alleviate the Hubble tension rather than explain the DESI observations, though the mechanism is closely related, and this naturally accommodates a possible self-interacting dark matter (SIDM) candidate as a by-product in the same way, as a consequence of constructing a gauge invariant Lagrangian. 

SIDM has been suggested as a possible resolution to tensions between $\Lambda$CDM and small-scale structure, for instance, the diversity of rotation curves observed in dwarf galaxies~\citep{Spergel_2000, Kamada_2017, Ren_2019}. $\Lambda$CDM (excluding baryonic physics) predicts NFW density profiles~\citep{Navarro_1996, Navarro_1997, Ludlow_2013}, though some dwarf galaxies are instead observed to have cored density profiles~\citep{Spergel_2000, Gentile_2004, Oman_2015, Santos_2017, Santos_2020}. Cored density profiles can be generated via baryonic feedback~\citep{Cores_Dwarf_Galaxies, Read_Gilmore_2005, Pontzen_2012,
Governato_et_al_2012, Teyssier_2013, Cintio_2013, Onorbe_2015, Chan_2015, Tollet_2016, Read_2016, Dutton_et_al_2019, 
Cores_Dwarf_Galaxies2, Lazar_2020, Jahn_2023, Azartash_2024}, which flattens the central density profile, though some dwarf galaxies appear to both have cored density profiles and contain too few stars for baryonic feedback to do this~\citep{Almeida_2024}. SIDM works to flatten the central density profile of the dark matter halo by incorporating non-gravitational interactions between particles, often through a Yukawa interaction, allowing them to scatter~\citep{Spergel_2000}. 
Our SIDM candidate is a heavy vector field that interacts via a scalar mediator. Heavy vectors interacting through a scalar mediator have previously been considered as an SIDM candidate~\citep{Duch_2018}, though our model is intriguing because the SIDM can naturally be incorporated through a minor extension of this framework, which already requires at least one U(1) symmetry. We find that our SIDM candidate has a velocity-dependent interaction cross section, which is necessary for the SIDM to generate cored halos in galaxies while also satisfying constraints from observations of galaxy clusters~\citep{Sagunski_2021, Correa_2021, Correa_2022}. 

\section{\label{sec:model}Our Model}

We have one complex scalar field, A, one real scalar field, B, and two dark Higgs fields, $\rm{H}_1$ and $\rm{H}_2$. We also include an ultralight real scalar, $\chi$, which could serve as dark energy. The model has a U(1) $\times$ U(1) gauge group, where B is uncharged. One might initially suspect that a scalar field will not cluster as dark matter because the kinetic terms could suggest a sound speed of $c$, though there is no such issue~\citep{Chen_2025}. 

Since the theory has a U(1) $\times$ U(1) gauge group, we must include two gauge bosons to allow for gauge invariance: $\alpha_\mu$ and $\beta_\mu$, where we define the Faraday tensors as $\Psi_{\mu \nu} \equiv \partial_\mu \Psi_\nu - \partial_\nu \Psi_\mu$, and the covariant derivative as $\mathcal{D}_\mu \equiv \partial_\mu - i(g_1 q_1 \alpha_\mu + g_2 q_2 \beta_\mu)$. We note that U(1) $\times$ U(1) is a subgroup of U(1) $\times$ U(1) $\times$ U(1), which can arise in string theory~\citep{Cvetic_2022}. We also note that anomalies will not appear in the model, as it does not include fermions~\citep{Feruglio_2021}. We can express our Lagrangian $\mathcal{L}$ as follows. For simplicity, we assume that the mixed quartic $(\mathrm{H}_1^\dagger \mathrm{H}_1) (\mathrm{H}_2^\dagger \mathrm{H}_2)$ has a small coupling constant, and can thus be neglected. 

\FloatBarrier

\begin{align}
\mathcal{L} \supset\;&
\mathcal{D}^\mu \mathrm{A}^{\dagger}\mathcal{D}_\mu \mathrm{A}
- m_\mathrm{A}^2 \mathrm{A}^{\dagger}\mathrm{A}
+ \frac{1}{2}\,\partial_\mu \mathrm{B}\,\partial^\mu \mathrm{B}
- \frac{1}{2}\, m^2(\chi)\mathrm{B}^2
\nonumber\\[0.4em]
&-\Big(g_{\mathrm{AB}}\, \mathrm{H}_2^{\dagger} \mathrm{A}\, \mathrm{B}^2
+ \text{h.c.}
\Big) -\frac{1}{4}\Big(
\alpha_{\mu\nu}\alpha^{\mu\nu}
+ \beta_{\mu\nu}\beta^{\mu\nu}
\Big)
\nonumber\\[0.4em]
&+\frac{1}{2}\partial_{\mu} \chi \partial^{\mu} \chi - \mathrm{V}(\chi) 
\nonumber\\[0.4em]
&+\sum_{i=1}^2
\Big[
\mathcal{D}^\mu \mathrm{H}_i^{\dagger}\mathcal{D}_\mu \mathrm{H}_i
+ m_{i}^2 \mathrm{H}_i^{\dagger}\mathrm{H}_i
- \lambda_{i}\big(\mathrm{H}_i^{\dagger}\mathrm{H}_i\big)^2
\Big] 
\end{align}

\FloatBarrier

To maintain generality, we have written the action, $\mathcal{L}$, with generic functions $m(\chi)$ and $\mathrm{V}(\chi)$, though in our analysis, we specify $m^2(\chi) = m_0^2 \exp(-2 \tilde c (\chi - \chi_0))$ and $\mathrm{V}(\chi) = \kappa_1 \exp(-b \chi) + \kappa_2 \exp(-c \chi)$. This potential was adopted from~\citep{Agrawal_2021}, where our function $m(\chi)$ matches that of~\citep{Agrawal_2021} at late times. We specify parameter values in the Methods section. Our model differs from theirs in that only a subcomponent of the dark matter couples to dark energy, where this subcomponent is produced via decays at late times. In principle, one could think of altering $m(\chi)$ and $\mathrm{V}(\chi)$ from the functions employed here, in which case this framework really represents a class of models, though we only consider one case here. One could also consider using $\mathrm{H_1}$ to trigger an earlier decay of some parent species to produce the field A when $\mathrm{H}_1$ gains a VEV, though we note that $\mathrm{H}_1$ is not required for alleviating the Hubble tension. 

Here we use \enquote{h.c.} to represent the Hermitian conjugate. The Lagrangian is invariant under local transformations of the form $\Phi \to e^{i q f(x)} \Phi$, where $q$ is the charge. If we instead had a global U(1) $\times$ U(1) symmetry, we would have transformations of the form $\Phi \to e^{i q f} \Phi$, and would not require gauge bosons, such as $\alpha_\mu$. We employ a local symmetry (rather than a global symmetry) because global symmetries are believed to be inconsistent with quantum gravity~\citep{Witten_2018, Harlow_2019, Harlow_2021}. For continuous global symmetries, such as U(1), one can motivate this from the black hole information paradox~\citep{Harlow_2021}, where one can roughly argue that a continuous global symmetry would suggest the existence of conserved charges that are destroyed by black hole evaporation. One can also motivate this from the Bekenstein-Hawking entropy formula via the argument found in the introduction of~\cite{Geng_2026}, though one should note that the global symmetries they construct later in their work are not applicable to astrophysical black holes, which do not possess a non-gravitational bath. Since we expect black hole evaporation to be unitary, continuous global symmetries should not exist in nature, which suggests that gauge symmetries are more natural. This motivates the inclusion of gauge bosons, which we shall see leads to interesting physical consequences, specifically a possible SIDM candidate, which has important consequences for small-scale structure. 

\begin{table}[h!]
\setlength{\tabcolsep}{3pt}        
\renewcommand{\arraystretch}{0.85} 
\small                              
\centering
\begin{tabular}{c|cc}
\hline
\textbf{Field} & \(\mathbf{U_1(1)}\) & \(\mathbf{U_2(1)}\) \\
\hline
\(\rm{A}\)   & 0 & 1 \\
\(\rm{H_1}\) & 1 & 0 \\
\(\rm{H_2}\) & 0 & 1 \\
\hline
\end{tabular}
\caption{Charge assignments ($q$) under $\rm{U}_1(1) \times \rm{U}_2(1)$.}
\end{table}

\FloatBarrier

When the first U(1) is spontaneously broken at early times, $\mathrm{H}_1$ gets a vacuum expectation value (VEV) and the gauge boson $\alpha_{\mu}$ gains a mass. When the second U(1) is spontaneously broken at late times, this generates a term of the form $g\mathrm{A B^2}$, which allows decays of the form $\mathrm{A} \to 2\mathrm{B}$. For a related mechanism that employs spontaneous symmetry breaking to activate decays, see~\citep{Mario_2019}, as well as~\citep{DESI_Cascade}. 

Here we use an interaction between dark matter and dark energy to allow for energy transfer from dark matter to dark energy, which allows for a larger $\mathrm{H}_0$ at late times~\cite{Agrawal_2021, Lucca_2020}. Since interactions between dark matter and dark energy are absent at early times in our construction, and only involve a subcomponent of the dark matter when active, this reduces fifth-force constraints on dark matter-dark energy interactions, and provides an alternative mechanism to that of~\citep{Agrawal_2021} for explaining why the coupling would suddenly emerge at late times. Evolving dark energy could additionally prevent the formation of a cosmological horizon at late times if it evolves appropriately, which would otherwise be problematic for quantum gravity~\citep{Dyson_2002}, and likely incompatible with M-theory~\citep{Bousso_2000}. 

We note that while one can in principle obtain late-time spontaneous symmetry breaking in the dark sector by carefully tuning the dark Higgs potential, it is perhaps desirable to use another mechanism. This could be accomplished by coupling the dark Higgs to the dark matter density, so that the redshifting matter density induces a sign change in the effective quadratic term, or alternatively through introducing a similar coupling between the dark Higgs and a slowly rolling ultralight scalar field. In either case, spontaneous symmetry breaking is dynamically induced at late times without requiring an extremely small transition temperature. For a related example where late-time spontaneous symmetry breaking is realized, see~\citep{MV_2026}.

\subsection{Self-Interacting Dark Matter Candidate}

\vspace{-0.4 cm}

When the first U(1) is spontaneously broken, giving the dark Higgs $\mathrm{H_1}$ and vector boson $\alpha_{\mu}$ masses, the kinetic term $\mathcal{D}^\mu \rm{H_1}^{\dagger} \mathcal{D}_\mu \rm{H_1}$ generates an interaction of the form $g h_1 \alpha_\mu \alpha^\mu$, which allows an $\alpha \alpha \to \alpha \alpha$ scattering process via a scalar mediator. For $m_{\alpha} \sim \rm{GeV}$ and $m_{h_1} \sim \rm{MeV}$, this provides a natural SIDM candidate, which could potentially resolve some of the challenges faced by $\Lambda$CDM, if $\alpha$ constitutes the majority of the dark matter. We display the $t$ and $u$-channel Feynman diagrams in Figure~\ref{fig:Diagrams}. We note that the $s$-channel is suppressed relative to the $t$ and $u$-channels in the relevant velocity regime ($v < \mathcal{O}(10^3$ km/sec)). See the Appendix for further details.

Our model does not strictly require that this SIDM candidate be included, though it can be very naturally embedded into our framework, especially if $\mathrm{H}_1$ is involved in triggering some earlier stage in the decay chain, which produces the field A through decays at early times. This feature is attractive, as SIDM can help to resolve important tensions involving small-scale structure, which motivates natural embeddings of SIDM candidates into astroparticle theory.

\FloatBarrier

\vspace{-0.3 cm}

\tikzset{
  vec/.style = {decorate, decoration={snake, amplitude=1.4pt, segment length=6pt}, line width=0.9pt},
  sca/.style = {dashed, line width=0.9pt},
  dot/.style = {circle, fill=black, inner sep=0pt, minimum size=3pt},
  lab/.style = {font=\small}
}

\begin{figure}[h]
\centering

\begin{tikzpicture}[line cap=round, line join=round, thick]
  \def\R{1.75}
  \def\dot{1.8pt}
  \usetikzlibrary{calc}
  \path
    ({\R*cos(45)},{\R*sin(45)})   coordinate (B2)
    ({\R*cos(135)},{\R*sin(135)}) coordinate (B3)
    ({\R*cos(-45)},{\R*sin(-45)}) coordinate (B1)
    ({\R*cos(-135)},{\R*sin(-135)}) coordinate (B4);

  \node[circle, fill=black, inner sep=\dot] (Vtop)    at (0,0.35*\R) {};
  \node[circle, fill=black, inner sep=\dot] (Vbottom) at (0,-0.35*\R) {};

  \draw[dotted] (Vtop) -- (Vbottom);
  \draw[vec] (Vtop) -- (B2);
  \draw[vec] (Vtop) -- (B3);
  \draw[vec] (Vbottom) -- (B1);
  \draw[vec] (Vbottom) -- (B4);
\end{tikzpicture}
\hfill
\begin{tikzpicture}[line cap=round, line join=round, thick]
  \def\R{1.75}
  \def\dot{1.8pt}
  \usetikzlibrary{calc}
  \path
    ({\R*cos(45)},{\R*sin(45)})   coordinate (B2)
    ({\R*cos(135)},{\R*sin(135)}) coordinate (B3)
    ({\R*cos(-45)},{\R*sin(-45)}) coordinate (B1)
    ({\R*cos(-135)},{\R*sin(-135)}) coordinate (B4);

  \node[circle, fill=black, inner sep=\dot] (Vtop)    at (0,0.35*\R) {};
  \node[circle, fill=black, inner sep=\dot] (Vbottom) at (0,-0.35*\R) {};

  \draw[dotted] (Vtop) -- (Vbottom);
  \draw[vec] (Vtop) -- (B1);
  \draw[vec] (Vtop) -- (B3);
  \draw[vec] (Vbottom) -- (B2);
  \draw[vec] (Vbottom) -- (B4);
\end{tikzpicture}

\caption{We provide Feynman diagrams for tree-level $\alpha\alpha \to \alpha\alpha$ scattering via a scalar mediator $h_1$. The $t$-channel diagram is on the left and the $u$-channel diagram is on the right.}
\label{fig:Diagrams}
\end{figure}

\FloatBarrier

\vspace{-1 cm}

\section{\label{sec:methods}Methods}

\vspace{-0.4 cm}

Our model allows for a decay pathway to open when a U(1) symmetry is spontaneously broken. We consider this in the context of the Hubble tension. Given rapid decays and a particular coupling to a quintessence field, we show that H(z) from our model can approximately map onto that of Agrawal et al., thus inheriting the fits to data~\citep{Agrawal_2021}. This equivalence only holds at the background level, and does not apply to perturbations. If we instead adopt a particular dynamical equation of state parameter for the daughter field, which past work by Naidoo et al. has used as a proxy for dark matter - dark energy interactions~\citep{Naidoo_Lambda_e_DM}, we find that we can similarly obtain approximate equivalence at the background level to Naidoo et al. While our decay pathway activated by spontaneous symmetry breaking can partially ground the phenomenological model of Naidoo et al. in field theory, it can only explain the sudden shift from $w_{\mathrm{dm}}=0$ to a dynamical $w_{\mathrm{dm}}$; it cannot automatically explain the particular dynamics of the equation of state parameter that Naidoo et al. employed, after $w_{\mathrm{dm}} \neq 0$~\citep{Naidoo_Lambda_e_DM}. Similar arguments apply for step-like changes in the matter equation of state parameter that have been used in other works in the literature, such as~\citep{Giani_2025}.

\subsection{Comparison to the Work of Agrawal et al.}

We numerically integrate the following system of differential equations from $z=1090$ to $z=0$ to compute the Hubble parameter, H(z). We use $\Theta ( t - t_i)$ to represent a step function that is nonzero after $\rm{U}_{i}(1)$ is spontaneously broken. One could consider replacing $\Theta ( t - t_i)$ with $f(z) = f_0 + \tanh(c_1z-c_2)$; we use $\Theta ( t - t_i)$ for simplicity. These coupled differential equations can be obtained through a modification of the equations given for generic $m(\chi)$ and $\mathrm{V}(\chi)$ for fermionic dark matter interacting with dark energy, provided in~\citep{Lopez_2010}. We have additionally verified that the equations of motion below are correct by deriving them, starting from our Lagrangian. Following the example of~\citep{Agrawal_2021}, we work in reduced Planck units, where the Planck mass is defined as $M_P = \sqrt{\hbar c / (8 \pi G)}$~\citep{Binetruy_2014}. This differs from the standard Planck mass by a factor of $1/\sqrt{8 \pi}$. 
\begin{equation}
  \dot{\rho}_{\mathrm{A}} + 3 \mathrm{H} \rho_{\mathrm{A}} = - \Gamma_{\mathrm{A}} \rho_{\mathrm{A}} \Theta(t - t_2)
\end{equation}
\vspace{-0.7 cm}
\begin{equation}
  \dot{\rho}_{\mathrm{B}} + 3 \mathrm{H} \rho_{\mathrm{B}} = -\tilde{c} \dot{\chi} \rho_{\mathrm{B}} + \Gamma_{\mathrm{A}} \rho_{\mathrm{A}} \Theta(t - t_2)
\end{equation}
\vspace{-0.7 cm}
\begin{equation}
  \ddot{\chi} + 3 \mathrm{H} \dot{\chi} - b \kappa_1 \exp(-b \chi) - c \kappa_2 \exp(-c \chi) = \tilde{c} \rho_{\mathrm{B}}
\end{equation}

The Friedmann equation takes the following form, where $\rho_\mathrm{A}$ and  $\rho_\mathrm{B}$ are the densities of the complex scalar and real scalar involved in the decay chain, whereas $\rho_{\mathrm{O}}$ is the total density of all other matter (baryons+other dark matter), $\rho_r$ is the radiation density, and $\rho_{\chi}$ is the dark energy density. We neglect any energy transfer to the massive radial mode ($h_2$) of a dark Higgs field $\mathrm{H}_2$ via three-body decays from an operator of the form $h_2 \mathrm{A} \mathrm{B}^2$; a more thorough analysis could incorporate this.  
\begin{equation}
\mathrm{H}^{2}=\frac{1}{3}
   (\rho_{\rm A}+\rho_{\rm B}
   +\rho_{\rm O}+\rho_{\rm r}+\rho_{\chi})
\label{eq:friedmann}
\end{equation}

We assume that decays of A $\to$ 2B begin at $z=15$, where A comprises 20$\%$ of the dark matter at recombination. For $z>15$ we assume dynamics of the quintessence field, $\chi$, are dictated by the steep component of the exponential potential~\citep{Agrawal_2021}. During this phase, we have the following approximation for $\rho_{\chi}$, where $w_{bd}$ is the effective equation of state parameter for the background, and $\rho_{bd}$ is the energy density for the background~\citep{Agrawal_2021}.

\begin{equation}
  \rho_{\chi} = \frac{3(1+w_{bd})}{b^2} \rho_{bd}
\end{equation}

At $z=15$, we begin solving the second order differential equation for $\chi$, where $\chi_0 = 0$ and $\dot{\chi}_0 \approx \sqrt{\frac{3}{b^2} \rho_{bd}}$. We set the energy densities at recombination by setting $h_i$ = 0.6825 for $\Lambda$CDM and 0.6912 for the case with dark matter - dark energy interactions, consistent with~\citep{Agrawal_2021}, where $\rm{H_{SI}}$ is $100 \rm{\frac{km}{sec Mpc}}$$h_i$ in units of $\rm{sec}^{-1}$. The radiation $\Omega_r$, baryonic matter $\Omega_b$, and dark matter $\Omega_{dm}$ components are determined by the following equations~\cite{Dodelson, Agrawal_2021}, which are consistent with Agrawal et al. We assume that a fraction $f$ of the dark matter lives in our cascade, and couples to dark energy. We later set $f=2/10$.

\begin{align}
\Omega_r &= 4.15 \times 10^{-5} /h_i^2 \\
\Omega_{dm} &= 0.1179/h_i^2
\\
\Omega_{b} &= 0.02237/h_i^2
\end{align}

\noindent The densities are then specified by the following equations, where $\rho_{m,i}$ and $\rho_{r,i}$ respectively refer to the total matter and radiation energy densities at recombination ($z=1090$). We convert the resulting values from SI to reduced Planck units before numerically integrating our system of differential equations. 

\vspace{-0.5 cm}

\begin{align}
\rho_{\rm crit,0} &= \frac{3 \mathrm{H_{SI}}^2}{8 \pi G} \\
\rho_{m,i} &= \Omega_m \rho_{\rm crit,0} (1 + z_{\rm rec})^3 \\
\rho_{r,i} &= \Omega_r \rho_{\rm crit,0} (1 + z_{\rm rec})^4 
\end{align}

\noindent Following~\citep{Agrawal_2021}, we use $c = 1/10$ and $b=50$. We also use $\kappa_1 = \kappa_2 = 0.81374 \times 10^{-120}$, $\Gamma_{\mathrm{A}} = 10^{-15} \hspace{0.1 cm} \mathrm{sec}^{-1}$, and $\tilde c = 1/2$. 

\subsection{Comparison to the Work of Naidoo et al.}

Naidoo et al. adopted the following function, $w_{\mathrm{dm}}(a)$, for the dark matter~\citep{Naidoo_Lambda_e_DM}, where the dark energy density was constant. They found that this can result in a larger $\mathrm{H}_0$ at late times, and could thus play a modest role in partially alleviating the Hubble tension. 

\begin{equation}
    w_{\mathrm{dm}}(a) = 
    \left\{
        \begin{array}{lr}
            w_{\mathrm{dm},0}\bigl(\frac{a-a_{\mathrm{nz}}}{1-a_{\mathrm{nz}}}\bigr), & \text{if } a \geq a_{\mathrm{nz}} \\
            0, & \text{if } a < a_{\mathrm{nz}} \\
        \end{array}
    \right.
\end{equation}

\noindent Naidoo et al. motivated this by arguing that it could arise as an effective description of dark matter - dark energy interactions, though they noted that this function for $w_{\mathrm{dm}}(a)$ is non-differentiable at the transition point~\citep{Naidoo_Lambda_e_DM}. This sudden shift from $w=0$ to $w \neq 0$ is difficult to motivate if the interaction between dark matter and dark energy is always active. It is also difficult to motivate if one has standard decays of A to B, where only B couples to dark energy, as the decay pathway does not switch on at a particular redshift, but instead is always active. Our model serves as one way to motivate this sudden shift~\cite{Naidoo_Lambda_e_DM} through a decay pathway that is activated via spontaneous symmetry breaking in the dark sector, while also naturally accommodating a self-interacting dark matter candidate. 

This only partially resolves the question of whether this phenomenological model can be grounded in particle theory; however, as this does not explain why the dark matter should have $w_{\mathrm{dm}}<0$ while the energy density of the dark energy remains constant. A realistic energy transfer between dark energy and dark matter would not do this, though one could argue that Naidoo et al. were simply using this as a rough phenomenological proxy, as dark matter-dark energy interactions can generate somewhat similar behavior at the background level, where we obtain a larger $\mathrm{H}_0$ at late times than in $\Lambda$CDM.

We shall see that the expansion history for our model maps onto that found from the model considered by Naidoo et al.~\cite{Naidoo_Lambda_e_DM} to a good approximation, given rapid decays, where field A has $w=0$, and field B has $w_{\mathrm{B}} = 2 w_0 \bigl( \frac{1}{1+z} - \frac{1}{2}\bigr)$. We emphasize that this approximate equivalence holds only at the background level, and does not carry over once perturbations are considered. Since detailed fits to data were conducted in this existing work~\citep{Naidoo_Lambda_e_DM}, it is redundant to repeat them here. 

We also note that a closely-related step-like change in $w$ for matter has already been considered by~\citep{Giani_2025}, where they conducted a fit to data, and found that DESI favors this simple model over $\Lambda$CDM at a similar significance to $w_0w_a$CDM, but without the problems implied by phantom dark energy. Interestingly,~\citep{Giani_2025} found the data favors $w = 0$ $\to$ $w \approx -1/6$, with a transition at $z \approx 1.1$. This is reminiscent of the type of construction considered here, where a decay pathway opens at a particular redshift, producing a subcomponent of dark matter that interacts with dark energy. Our work is distinct from ~\citep{Naidoo_Lambda_e_DM, Giani_2025} for at least making partial progress towards providing a specific microphysical mechanism, and also for naturally incorporating an SIDM candidate as a by-product. We note that~\citep{Giani_2025} were not seeking to alleviate the Hubble tension. For further related work showing a preference for $w_{\rm{dm}}<0$ at late times, see~\citep{Yao_2025, Xu_2026}.

In the equations below, $\rho_\mathrm{A}$ and  $\rho_\mathrm{B}$ are the densities of the complex scalar and real scalar involved in the decay chain, whereas $\rho_{\mathrm{O}}$ is the total density of all other matter (baryons+other dark matter), $\rho_r$ is the radiation density, and $\rho_{\Lambda}$ is the dark energy density. We neglect any energy transfer to the massive radial mode ($h_2$) of a dark Higgs field $\mathrm{H}_2$ via three-body decays from an operator of the form $h_2 \mathrm{A} \mathrm{B}^2$; a more thorough analysis could incorporate this. 
\begin{align}
\dot{\rho}_{\rm A}+3\mathrm{H}(1+w_{\rm A})\rho_{\rm A} &= -\Theta(t-t_2)\Gamma_{\rm A}\rho_{\rm A} \\
\dot{\rho}_{\rm B}+3\mathrm{H}(1+w_{\rm B})\rho_{\rm B} &= \Theta(t-t_2)\Gamma_{\rm A}\rho_{\rm A} 
\end{align}

\vspace{-0.8 cm}

\begin{equation}
\mathrm{H}^{2}=\frac{8\pi G}{3}
   (\rho_{\rm A}+\rho_{\rm B}
   +\rho_{\rm O}+\rho_{\rm r}+\rho_{\Lambda})
\label{eq:friedmann}
\end{equation}

We integrate this system of equations from recombination at $z_{\rm rec}=1090$ to $z=0$, where we set the initial conditions for the system in much the same way as when we compared our framework to the work of Agrawal et al. We employ the quantities below to set the initial conditions for $\Lambda$CDM and the model of Naidoo et al.~\citep{Naidoo_Lambda_e_DM}. 

\begin{table}[h!]
\setlength{\tabcolsep}{3pt}        
\centering
\begin{tabular}{c|ccc}
\hline
\textbf{Model} & \(\mathbf{\Omega_{dm}h^2}\) & \(\mathbf{\Omega_{b}h^2}\) 
& \(\mathbf{h}\) \\
\hline
\(\Lambda\mathrm{CDM}\)   & 0.1193 & 0.02245 & 0.6799\\
Naidoo & 0.1185 & 0.02251 & 0.688 \\
\hline
\end{tabular}
\caption{Parameters used in setting initial conditions for $\Lambda\mathrm{CDM}$ and the model of Naidoo et al.~\citep{Naidoo_Lambda_e_DM} }
\end{table}

\FloatBarrier

We have $f = \rho_{\mathrm{A}}/\rho_{\mathrm{DarkMatter}}$ at $z=1090$ as a free parameter that sets the fraction of dark matter living in the cascade at recombination. We also have the decay rate, $\Gamma_{\mathrm{A}}$, and the redshift at symmetry breaking, $z_{\mathrm{BREAK}}$, as free parameters. Note that $z_{\mathrm{BREAK}}$ corresponds to the cosmic time $t_2$. 

\section{\label{sec:results_discussion}Results and Discussion}

\subsection {Effect on H(z)}

\begin{figure}[h!]\includegraphics[width=2.5 in, height=2.0 in]{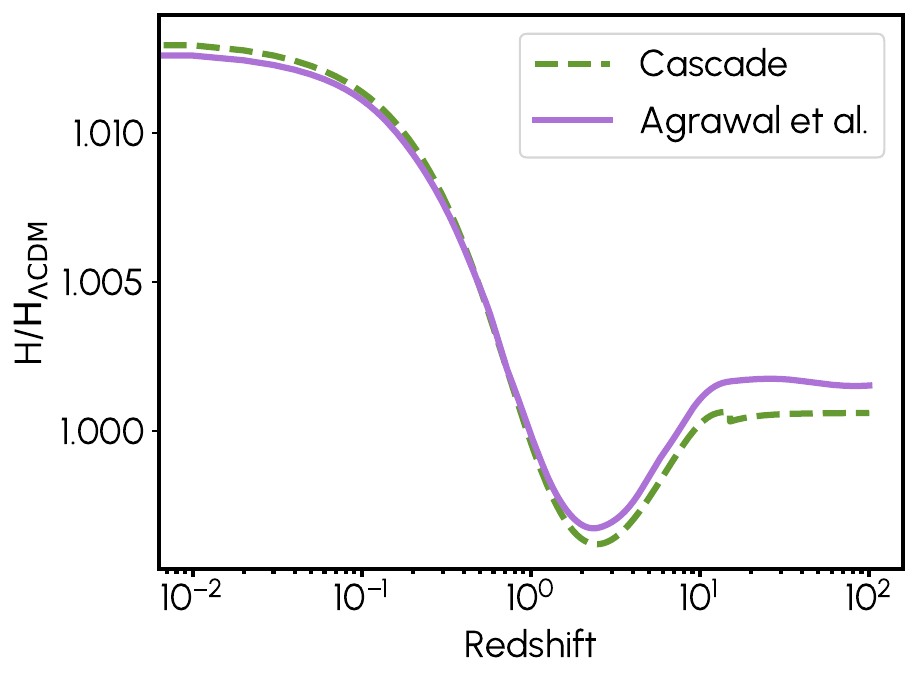} 
\caption{We plot the ratio of the Hubble parameter from our model to that obtained from $\Lambda$CDM. We do the same for the model considered by Agrawal et al.~\cite{Agrawal_2021}, as shown in Figure 2 of their paper. We see that the expansion histories are nearly identical.}
\label{fig:H_H_Agrawal}
\end{figure}

\begin{figure}[h!]\includegraphics[width=2.5 in, height=2.0 in]{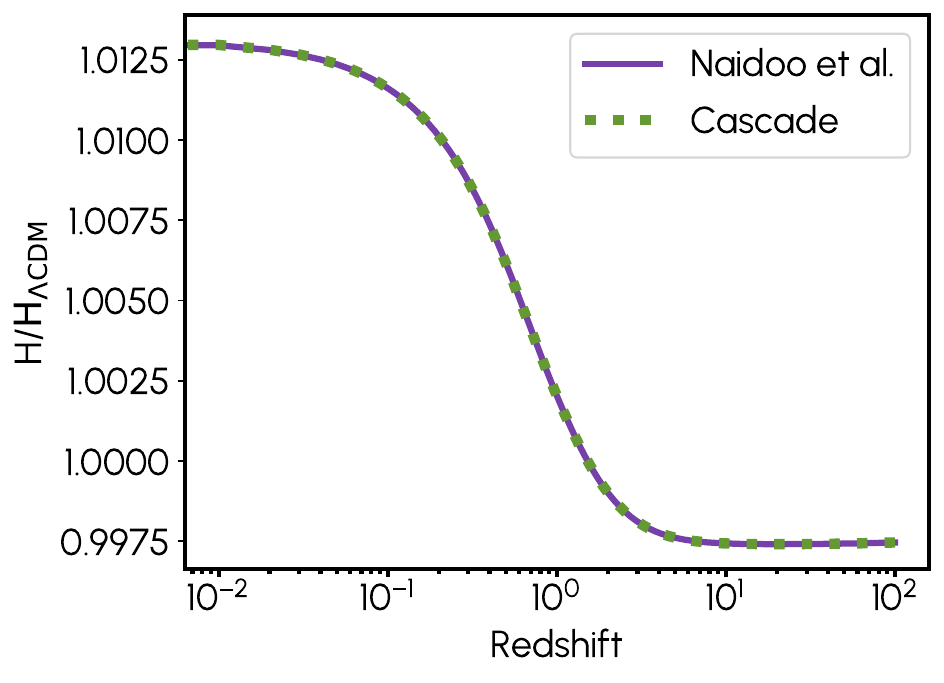} 
\caption{We plot the ratio of the Hubble parameter from our model to that obtained from $\Lambda$CDM. We do the same for the model considered by Naidoo et al.~\cite{Naidoo_Lambda_e_DM}, where we set $w_0$ in the model of Naidoo et al. to $-1/100$. In our model, we set $w_0 = -4/100$, $z_{\mathrm{BREAK}}=1$, $\Gamma_{\mathrm{A}} = 10^{-16} \hspace{0.1 cm} \mathrm{sec}^{-1}$, and $f = 1/4$. We see that the expansion histories are nearly identical.}
\label{fig:H_H_Naidoo}
\end{figure}

The code used in the following analysis is publicly available~\citep{Hoelscher2025Code, Two_Stage_Code}. We illustrate in Figure~\ref{fig:H_H_Agrawal} that the expansion history from our model maps onto that from the model considered by Agrawal et al.~\citep{Agrawal_2021} to a good approximation, when considering interactions between quintessence and the daughter species, with $f=2/10$. We see something similar when comparing to Naidoo et al.~\citep{Naidoo_Lambda_e_DM} in Figure~\ref{fig:H_H_Naidoo}, where the daughter species is given a dynamical equation of state of the form used by Naidoo et al., rather than explicitly coupling it to dark energy. We set $w_0$ in the model of Naidoo et al. to $-1/100$. In our model, we set $w_0 = -4/100$, $z_{\mathrm{BREAK}}=1$, $\Gamma_{\mathrm{A}} = 10^{-16} \hspace{0.1 cm} \mathrm{sec}^{-1}$, and $f = 1/4$. We also briefly mention that Braglia, Chen, and Loeb argue that existing constraints from structure do not strongly constrain these types of models, as the evolution in $w$ occurs at late times~\citep{Braglia_2025}.

\FloatBarrier
\FloatBarrier

\FloatBarrier
\FloatBarrier

\subsection{SIDM Cross Section}

\begin{figure}[H]
    \centerline{
    \includegraphics[width=2.5 in, height=2.0 in]{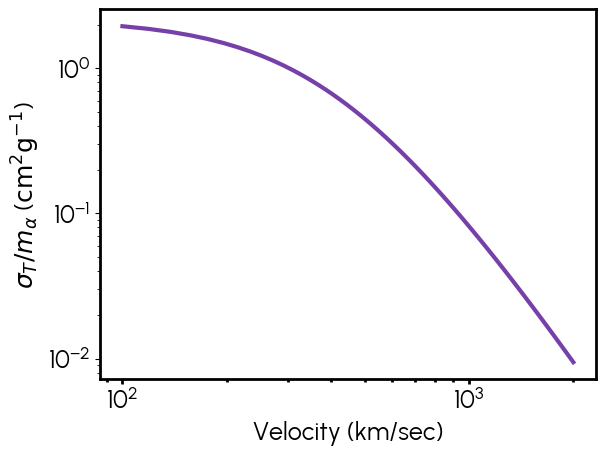}}
    \caption{We plot the tree-level momentum-transfer cross section for the scattering process $\alpha \alpha \to \alpha \alpha$ with $m_\alpha$ = 1 GeV and $m_{h_1} = 0.001$ GeV. One can see that this decreases with increasing relative velocity, potentially enabling the SIDM to evade constraints from galaxy clusters while still producing cored density profiles in galaxies. } 
\label{fig:Momentum_Transfer}
\end{figure}

\begin{figure}[H]
    \centerline{
    \includegraphics[width=2.5 in, height=2.0 in]{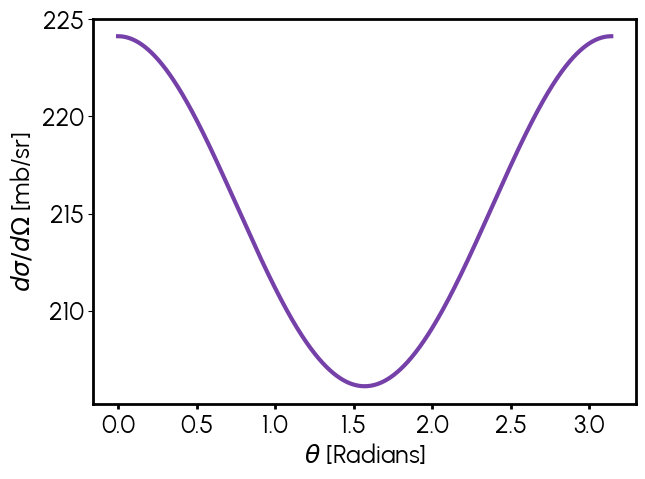}}
    \caption{We plot the tree-level differential cross section for the scattering process $\alpha \alpha \to \alpha \alpha$ with $m_\alpha$ = 1 GeV, $m_{h_1} = 0.001$ GeV, and a relative velocity of 200 km/sec. This velocity is relevant for the Milky Way scale. } 
\label{fig:Differential_Cross_Sect_200}
\end{figure}

\begin{figure}[h!]
    \centerline{
    \includegraphics[width=2.5 in, height=2.0 in]{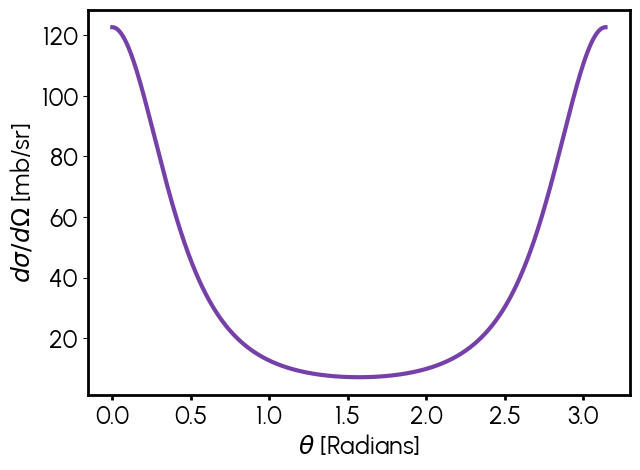}}
    \caption{We plot the tree-level differential cross section for the scattering process $\alpha \alpha \to \alpha \alpha$ with $m_\alpha$ = 1 GeV, $m_{h_1} = 0.001$ GeV, and a relative velocity of $10^3$ km/sec, which is relevant for the galaxy cluster scale. One can see that this cross section is suppressed as compared to the result for 200 km/sec.} 
\label{fig:Differential_Cross_Sect_1000}
\end{figure}

Our computation of the tree-level SIDM cross section is publicly available in~\citep{Hoelscher2025Code}. We plot the momentum-transfer cross section as a function of relative velocity in Figure~\ref{fig:Momentum_Transfer}, with $m_{\alpha} =$ 1 $\rm{GeV}$, $m_{h_1} =$ 0.001 $\rm{GeV}$, and $g = 0.0175$. These are free parameters of the model, so we are free to set them to values that produce scattering that is consistent with observational constraints~\cite{Correa_2021}. In the non-perturbative regime, one would need to instead use partial wave analysis to compute the cross section numerically. Note that we use $m_{\alpha}$ for the mass of the vector, $m_{h_1}$ for the mass of the scalar mediator, and $g$ to represent the effective coupling constant. We also show the differential scattering cross section in Figure~\ref{fig:Differential_Cross_Sect_200} and Figure~\ref{fig:Differential_Cross_Sect_1000}. We see that the cross section decreases as relative velocity increases. For our parameter values, we have $(g^2 / 4 \pi)(1 / m_\alpha m_{h_1}) \approx 0.024 << 1$, so perturbation theory is appropriate as an approximation. 

We define the momentum-transfer cross section, $\sigma_T$,~\citep{Correa_2022} below. The momentum transfer cross section is a more useful quantity for SIDM than the total cross section, as it down-weights forward and backward scattering by including $(1-|\rm{cos (\theta)}|)$ within the integral. We use $\theta$ for the scattering angle, $\Omega$ for the solid angle, $\mathrm{E}_{\mathrm{cm}}$ for the energy in the center-of-mass frame, and $\mathcal{M}$ for the S-matrix element. We include a closed-form expression for $|\mathcal{M}|^2$ in the Appendix. This was produced from the $t$ and $u$-channel Feynman diagrams (see Figure~\ref{fig:Diagrams}).
\begin{equation}
    \sigma_{T} = 2 \int d\Omega(1-|\rm{cos (\theta)}|) \frac{|\mathcal{M}|^2}{64 \pi^2 \rm{E_{cm}}^2}
\end{equation}

\FloatBarrier

\vspace{-0.5 cm}

\FloatBarrier

\FloatBarrier

\FloatBarrier

\section{\label{sec:conclusion}Conclusion}

We see that we can construct a simpler generalization of our framework in~\citep{DESI_Cascade} to perhaps partially alleviate the Hubble tension while also incorporating a natural self-interacting dark matter candidate as a by-product of constructing a gauge invariant theory. Our framework naturally incorporates an SIDM candidate with a velocity-dependent scattering cross section, which could potentially allow it to produce cored dark matter density profiles at the galaxy scale, while avoiding constraints from galaxy clusters. Future work can consider production mechanisms for the dark matter, which would be relevant for specifying acceptable ranges for particle masses. Future work could also consider constraints from~\citep{Huang_2024}, or could study combining such a model with new physics in the early Universe that modifies the sound horizon. The resolution to the Hubble tension may then be a combination of early and late-time new physics, as well as systematics, e.g. see~\citep{Vagnozzi_2023}, which has argued that both early and late-time modifications to $\Lambda$CDM may be required. One could also consider modifying the potential, $\mathrm{V}(\chi)$, as well as the coupling, $m(\chi)$.

\vspace{-0.75 cm}

\section{Appendix}

\subsection{Justification for Neglecting the $s$-Channel for SIDM}

\noindent We define the Mandelstam variables as shown below:
\begin{equation}
    s = (2\mathrm{E})^2 = \mathrm{E}^2_{\mathrm{CM}},
\end{equation}

\vspace{-1.1 cm}

\begin{equation}
    t = -2p^2(1-\cos(\theta)),
\end{equation}

\vspace{-1.1 cm}

\begin{equation}
    u = -2p^2(1+\cos(\theta)).
\end{equation}

\vspace{0.25 cm}

\noindent For $v <<< 1$, $\mathrm{E_{\mathrm{CM}}^2} \approx 4 m_{\alpha}^2$ and $p \approx m_{\alpha} v$. Note that $m_{\alpha} >>> m_{h_1}$. This suggests:
\begin{equation}
    \biggl| \frac{1}{s - m_{h_1}^2} \biggr| <<< \biggl| \frac{1}{t - m_{h_1}^2} \biggr|,
\end{equation}
\begin{equation}
    \biggl| \frac{1}{s - m_{h_1}^2} \biggr| <<< \biggl| \frac{1}{u - m_{h_1}^2} \biggr|.
\end{equation}

\noindent While the $s$-channel exists, it is suppressed in the velocity regime we care about for dark matter halos (relative velocities less than $ \mathcal{O}$ ($10^3$) km/sec). This then justifies working with only the $t$ and $u$-channel diagrams, when computing a tree-level cross section. 

\subsection{S-Matrix Amplitude for Tree-Level Scattering Cross Section}

Below we provide a closed-form expression for the S-matrix amplitude we have used in plotting the SIDM cross section. Here, the relevant interaction operator is $g h_1 \alpha_{\mu} \alpha^{\mu}$. Our calculation is available as a PDF in~\citep{Hoelscher2025Code}. This was manually computed from the $t$ and $u$-channel Feynman diagrams. For brevity, we define $C_1$, $C_2$, and $C_3$ as follows. Here, $p$ represents the momentum, $\mathrm{E_{cm}}$ represents the energy in the center-of-mass frame, $\theta$ is the scattering angle, $m_{h_1}$ is the mass of the scalar mediator, and $m_{\alpha}$ is the mass of the vector field, $\alpha_{\mu}$. 

\begin{align}
C_1 &= \tfrac14 \mathrm{E_{cm}}^2 - p^2 \cos(\theta) \\
C_2 &= \tfrac14 \mathrm{E_{cm}}^2 + p^2 \cos(\theta) \\
C_3 &= \tfrac14 \mathrm{E_{cm}}^2 + p^2 
\end{align}

\FloatBarrier

\begin{equation}
\begin{aligned}
|\mathrm{\mathcal{M}}|^2 &= 
\frac{8 g^4}{9 m_\alpha^8}
\Bigg[
  \frac{
    4 m_\alpha^4 (C_1^2 + C_2^2 + C_3^2)
    + 2 C_1^2 C_2^2
    - 8 C_1 C_2 C_3 m_\alpha^2
  }{
    (m_{h_1}^2 + 2p^2)^2 - 4p^4\cos^2(\theta)
  }  \\[3pt]
&\qquad\qquad
 + \frac{(C_1^2 + 2 m_\alpha^4)^2}{
    (m_{h_1}^2 - 2p^2\cos(\theta) + 2p^2)^2
  } \\[3pt]
 &\qquad\qquad + \frac{(C_2^2 + 2 m_\alpha^4)^2}{
    (m_{h_1}^2 + 2p^2\cos(\theta) + 2p^2)^2
  }
\Bigg]
\label{eq:M2}
\end{aligned}
\end{equation}

\bibliography{apssamp}

\end{document}